\documentclass[11pt,a4paper]{article}

\usepackage[utf8]{inputenc}
\usepackage[T1]{fontenc}
\usepackage{amsmath,amsfonts,amssymb}
\usepackage{amsthm}
\usepackage{graphicx}
\usepackage{booktabs}
\usepackage{multirow}
\usepackage[table]{xcolor}
\usepackage{algorithm}
\usepackage{algorithmic}
\usepackage{array}
\usepackage{textcomp}
\usepackage{stfloats}
\usepackage{url}
\usepackage{verbatim}
\usepackage{hyperref}
\usepackage{cite}
\usepackage[margin=1in]{geometry}
\usepackage{authblk}

\newtheorem{definition}{Definition}
\newtheorem{property}{Property}

\begin{document}

\title{Blind PRNG Hijacking: An Undetectable\\
Integrity-Preserving Attack Against LLM Watermarking}

\author[1]{Ziyang You}
\author[1]{Huilong He}
\author[2]{Xiaoke Yang}
\author[1,3]{Xuxing Lu\thanks{Corresponding author: Xuxing Lu (e-mail: xuxinglu@um.edu.mo).}}
\affil[1]{Fujian Provincial Key Laboratory of Automotive Electronics and Electric Drive, School of Electronic, Electrical and Physics, Fujian University of Technology, Fuzhou 350118, China}
\affil[2]{School of Humanities, Fujian University of Technology, Fuzhou 350118, China}
\affil[3]{Institute of Applied Physics and Materials Engineering, University of Macau, Macau 999078, China}

\date{}

\maketitle

\begin{abstract}
Cryptographic watermarking is a leading defense for attributing text generated
by large language models (LLMs). Existing schemes, including KGW, Unigram, and
DipMark, derive their security guarantees from the assumption that the
underlying pseudo-random number generator (PRNG) is trustworthy. This work
introduces \emph{SeedHijack}, the first supply-chain attack on LLM watermarking
that is simultaneously (i)~\emph{blind}---requiring no knowledge of the
watermark key, detector, or model logits, (ii)~\emph{integrity-preserving}---amplifying
rather than erasing the watermark signal, and (iii)~\emph{orthogonal} to
detection---the attack-induced bias is statistically independent of all
content-side detector statistics, ensuring that amplification and evasion coexist
without trade-off. Rather than perturbing generated text, SeedHijack replaces
the PRNG at the supply-chain layer, biasing green-list selection without
altering output tokens or degrading text quality. Across three watermarking
schemes and three open-source LLMs, the attack triggers \emph{0/6}
state-of-the-art \emph{content-side} statistical detectors while inflating the watermark $z$-score
up to $2.42\times$ (system-level defenses such as entropy-source attestation remain orthogonal and complementary). A quantum random number generator (QRNG) countermeasure is
shown to fully neutralize the attack while preserving benign watermarking
utility. These findings establish PRNG integrity as a first-class security
requirement for cryptographic content-provenance systems.
\end{abstract}


\bigskip

\section{Introduction}
The rapid deployment of large language models (LLMs) in
content production has made provenance attribution a pressing concern
for platforms, regulators, and end users. Cryptographic watermarking embeds an
imperceptible statistical signal into model outputs and has become the dominant
technical answer~\cite{kirchenbauer2023,zhao2023unigram,wu2023dipmark}.
Schemes such as KGW, Unigram, and DipMark all share a common architectural
foundation: a hash-keyed pseudo-random number generator (PRNG) partitions the
vocabulary into a ``green'' and a ``red'' list, and the generator is biased
toward sampling green tokens. The detector replays the same PRNG to recover the
green list and applies a one-sided statistical test. Crucially, the security
arguments behind these schemes implicitly assume that the PRNG itself is
trustworthy.

Software supply-chain attacks, however, have escalated from theoretical risks to verified incidents affecting critical infrastructure, with SolarWinds~\cite{peisert2021solarwinds} and the xz Utils backdoor~\cite{jia2024xzbackdoor} demonstrating that adversaries can quietly subvert widely trusted dependencies that millions of downstream systems rely upon. The machine-learning ecosystem is particularly exposed: modern training and inference pipelines depend on thousands of third-party libraries, many of which handle cryptographic primitives and random number generation without independent auditing. Within these pipelines, randomness itself is sourced from layered software-supply-chain components (system entropy pools, hash functions, and seeding utilities), each of which is a plausible attack surface. We identify the pseudorandom number generator (PRNG), a ubiquitous but under-scrutinized dependency in every watermarking deployment, as a novel supply-chain attack surface whose compromise enables integrity-preserving manipulation of LLM watermarking systems.

Existing attacks on LLM watermarks consistently trade off two properties that a
realistic adversary would like to obtain simultaneously: stealth (the
attack leaves no statistical fingerprint) and watermark preservation
(the manipulated output continues to be attributed to the targeted source).
Paraphrasing attacks~\cite{krishna2023paraphrase} can suppress detection but
fundamentally destroy the watermark, defeating their stealth value when the
attacker's goal is misattribution rather than removal.
Token-editing and substitution attacks~\cite{kirchenbauer2024reliability,wu2024bypassing} are detectable by perplexity-, Kolmogorov--Smirnov
(KS)-, and burstiness-based statistical tests, and watermark-stealing attacks~\cite{jovanovic2024watermark,zhang2024stealing} that reconstruct the keyed green-list partition from observed outputs typically require large query budgets and leave entropy-side traces. Prompt-injection attacks~\cite{jovanovic2024prompt} are only
partially effective and remain visible to behavioral monitors.
Recent analyses of inherent design trade-offs in watermark construction~\cite{pang2024nofree} further confirm that existing schemes cannot jointly optimize robustness, detectability, and text quality.
Even zero-shot detectors based on probability curvature~\cite{mitchell2023detectgpt} can flag crude manipulations.
To date, no
published attack achieves the conjunction of undetectability across a
multi-detector suite and strict integrity of the watermark signal.

Recent work has shown that PRNG manipulation can force exact token injection in LLM sampling without altering model logits~\cite{you2026seedhijack}, establishing the feasibility of supply-chain attacks at the randomness layer. However, token injection and watermark manipulation pose fundamentally different challenges: the latter requires the attacker to navigate the interplay between PRNG states and watermark-keyed partitions, to preserve rather than disrupt the embedded signal, and to survive scrutiny by statistical detectors specifically designed to flag distribution anomalies.

This paper addresses these challenges. By relocating the attack from the output layer to the seed layer and exploiting a newly identified green-list orthogonality property, the proposed attack simultaneously amplifies the watermark $z$-score and remains \emph{content-side} undetectable across a six-detector suite. The contributions are summarized as follows.
\begin{itemize}
\item A \emph{blind} attack mode that requires no access to the watermark
detector, the green-list seed, or the model logits, broadening the realistic
threat model to opaque deployment environments where prior attacks degenerate.
\item The first \emph{Integrity-Preserving Attack} paradigm against LLM
watermarking, in which the attacker leaves the watermark verifiable (and indeed amplifies its $z$-score) while remaining undetectable by content-side
analysis, breaking the long-standing stealth-versus-preservation trade-off.
\item A \emph{green-list orthogonality} property: the attacker's biased target
set $T$ and the watermark's keyed green list $G_t$ are cryptographically
independent, so attack-induced bias is provably non-leaking into any
content-side detector statistic. This property is the theoretical pillar that
simultaneously justifies undetectability and amplification.
\item A QRNG-based defense rooted in physical entropy, with a detailed
discussion of why software-only countermeasures are structurally insufficient.
\end{itemize}

Empirically, the proposed attack triggers $0$ out of $6$ state-of-the-art content-side stealth detectors, inflates the watermark $z$-score by up to $2.42\times$, and
generalizes across three watermarking schemes and three open-source LLMs;
the QRNG-based defense fully neutralizes the attack with no degradation to
benign watermarking utility.

The remainder of this paper is organized as follows.
Section~\ref{sec:related} surveys related work.
Section~\ref{sec:threat} formalizes the threat model and problem.
Section~\ref{sec:method} presents the SeedHijack methodology.
Section~\ref{sec:eval} reports the experimental evaluation.
Section~\ref{sec:defense} analyzes defenses, Section~\ref{sec:disc} discusses
implications and limitations, and Section~\ref{sec:concl} concludes.

\section{Related Work}\label{sec:related}

\subsection{LLM Watermarking Schemes}
Modern LLM watermarking embeds a hash-keyed statistical bias into token
sampling so that a verifier holding the same key can later detect generated
text. The seminal scheme of Kirchenbauer~\textit{et al.}~\cite{kirchenbauer2023}
(``KGW'') partitions the vocabulary into a green and a red list at every step
using a hash of recent context, then adds a positive logit bias $\delta$ to
green tokens. Detection uses a one-sided $z$-test on the proportion of green
tokens. Variants such as the soft watermark~\cite{kirchenbauer2024reliability}
and Unigram~\cite{zhao2023unigram} replace the context-dependent green list
with a global, context-free key; this makes the detector simpler and more
robust to small edits at the cost of slightly lower per-token entropy.
Distortion-free schemes such as
DipMark~\cite{wu2023dipmark,kuditipudi2023robust} preserve the original output
distribution in expectation, ensuring that watermarked text is statistically
indistinguishable from un-watermarked text under appropriate metrics.
Undetectable watermarks~\cite{christ2024undetectable} achieve cryptographic undetectability under computational assumptions; semantic watermarks~\cite{hou2023semstamp} offer an alternative embedding strategy. Fernandez~\textit{et al.}~\cite{fernandez2023three} consolidate design principles shared across these methods.
A parallel line of robustness-oriented defenses, including semantic-invariant watermarks~\cite{liu2024semantic}, context-aware green-red list construction~\cite{guo2024context}, encoder-decoder watermarking frameworks~\cite{zhang2024remark}, and resilience-enhancing schemes against scrubbing and spoofing~\cite{shen2025enhancing}, has hardened watermarks against output-layer perturbations.
All such
schemes share one common assumption: the keyed PRNG that generates the green
list is honestly executed both at generation and at verification time.

\subsection{Attacks on LLM Watermarks}
the literature on attacks against LLM watermarks operates almost exclusively at
the output layer, modifying the generated text after sampling.
Paraphrasing attacks feed
watermarked outputs through a paraphraser to disrupt the green/red token
distribution; they are highly effective at suppressing detection but destroy
the watermark, which is the very signal a misattribution adversary would
prefer to preserve. Token-editing and synonym-substitution
attacks~\cite{kirchenbauer2024reliability,zhang2024watermark} replace a small
fraction of tokens but leave detectable traces in $\mathrm{perplexity}$, burstiness, and
$n$-gram statistics.
Prompt-injection
attacks~\cite{jovanovic2024prompt,liu2024survey} attempt to coerce the model
into producing un-watermarked text but are partial in effect and visible to
behavioral monitors. A common limitation runs through all of these
approaches: they cannot simultaneously satisfy stealth (no statistical
fingerprint under a multi-detector suite) and watermark integrity (the
embedded signal continues to verify under the original key). The attack
proposed in this work bypasses this trade-off by operating on the PRNG
before sampling, leaving the output text statistically identical to a
benign generation.

\subsection{PRNG Security in ML Systems}
Supply-chain attacks against ML pipelines have been documented at the level of
training data~\cite{carlini2024poisoning}, model
weights~\cite{shafahi2018poison}, and software
dependencies~\cite{zhang2023supplychain}.
High-profile incidents such as SolarWinds~\cite{peisert2021solarwinds} and the xz Utils backdoor~\cite{jia2024xzbackdoor} demonstrate that supply-chain insertion of cryptographic backdoors is a practical threat; analogous risks have also been identified in ML model repositories.
The randomness layer, however, has
received comparatively little attention, despite being a low-cost and
high-leverage point of compromise. Predictability of software PRNGs has been
exploited in cryptographic
contexts~\cite{dorrendorf2009cryptanalysis,heninger2012mining}, and Dual~EC
remains the canonical example of a backdoored generator deployed at scale.
Checkoway~\textit{et al.}~\cite{checkoway2014dualec} provided practical exploitation of Dual~EC in TLS, confirming that PRNG backdoors translate directly to protocol-level compromise.
In ML, biased randomness has been studied as a fairness or correctness
issue but, to the authors' knowledge,
has not been mounted as a deliberate attack vector against
cryptographically anchored content-provenance schemes. This work is the
first to formalize PRNG seed hijacking as an attack surface against LLM
watermarking and to demonstrate, both theoretically and empirically, that the
resulting attack is integrity-preserving and undetectable.

\subsection{Seed Hijacking}
You~\textit{et al.}~\cite{you2026seedhijack} first demonstrated that manipulating the PRNG output in LLM sampling pipelines enables exact token injection without modifying model logits or post-processing. Their work achieved 99.6\% injection accuracy on GPT-2 (124M) across nine sampling configurations, scaling to 100\% on four aligned models (1.5B--7B) spanning RLHF, SFT, and reasoning distillation. A hardware QRNG defense was shown to neutralize the attack with negligible overhead (+0.6\% latency).

The present work departs from~\cite{you2026seedhijack} in both threat objective and technical scope. Where the original attack targets arbitrary token injection, the attack proposed here targets watermark integrity: amplifying and preserving a legitimate watermark signal while maintaining content-side undetectability. This shift introduces three challenges absent from the injection setting: (i)~the attacker must coordinate PRNG manipulation with the unknown or partially known green-list partition (formalized in the blind mode); (ii)~the output must remain undetectable under a multi-detector statistical suite rather than passing a single behavioral check; and (iii)~the attack must strengthen the watermark rather than suppress it, requiring a formal analysis of the orthogonality between PRNG manipulation and watermark bias. These extensions motivate the three security definitions (Definitions~\ref{def:preserve}--\ref{def:stealth}), the cross-scheme generalization across KGW, Unigram, and DipMark, and the quantitative defense validation reported in Section~\ref{sec:defense}.

\section{Threat Model and Problem Formulation}\label{sec:threat}
\begin{figure*}[!t]
\centering
\includegraphics[width=\linewidth]{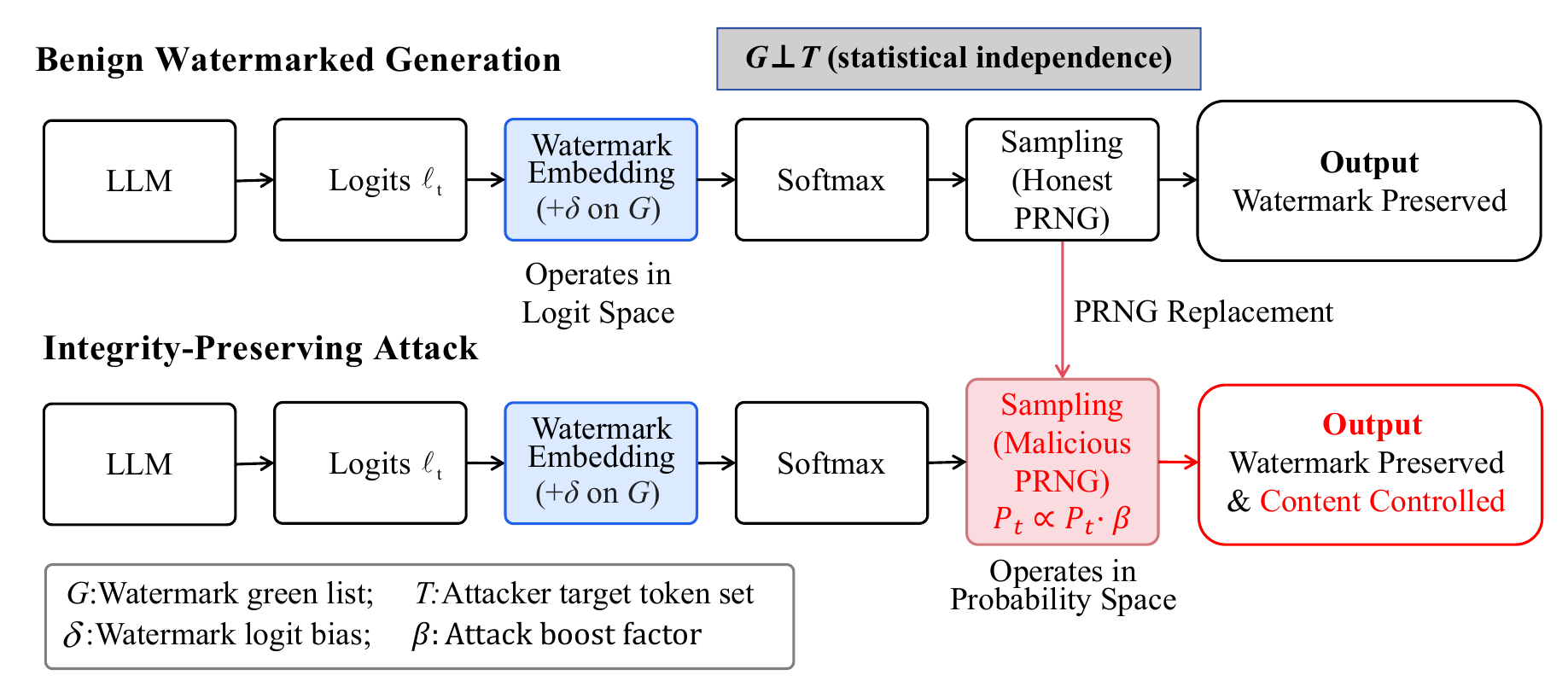}
\caption{Dual-flow comparison of watermarked LLM inference. \textbf{Top}:
benign watermarked generation where the watermark adds bias $+\delta$ to
green-list tokens $G$ in logit space. \textbf{Bottom}: SeedHijack
attack where a malicious PRNG replaces the honest one at the supply-chain
layer, biasing sampling toward a target set $T$ in probability space.
Because $G$ and $T$ are statistically independent (green-list
orthogonality), the watermark $z$-score is preserved while the attacker
gains content control.}
\label{fig:architecture}
\end{figure*}

This section formalizes the operating environment, the adversary's
capabilities, and the precise notion of attack success that this work
targets. The aim is to characterize a class of attacks that watermark
designers have so far implicitly assumed away by trusting the underlying
randomness source, and to state the security requirements an attack must
fulfill to qualify as integrity-preserving and undetectable.
A high-level view of the LLM inference pipeline together with the
proposed PRNG-layer attack-injection point is shown in
Fig.~\ref{fig:architecture}.

\subsection{System Model}\label{ssec:sysmodel}
A modern LLM inference pipeline produces tokens $y_t$ autoregressively from
logits $\ell_t \in \mathbb{R}^{|V|}$ over a vocabulary $V$. A cryptographic
watermarking scheme $\mathcal{W}$ is parameterized by a secret key $k$ and
at each step performs three operations: (i) a keyed PRNG $\mathcal{R}_k$ is
seeded with context-derived state $s_t = h_k(y_{<t})$ to draw a binary
partition $G_t \subseteq V$ (the green list); (ii) a bias function
$\phi(\ell_t, G_t)$ shifts mass toward $G_t$ (e.g., logit shift $+\delta$
for KGW, multiplicative reweighting for
DipMark, or a global green list for
Unigram); (iii) the modified distribution
$p_t = \mathrm{softmax}(\phi(\ell_t, G_t))$ is sampled to obtain $y_t$.
A detector $\mathcal{D}_{\mathcal{W}}$ replays $\mathcal{R}_k$ on the
published output $y_{1:n}$ to reconstruct $\{G_t\}_{t=1}^{n}$ and applies a
one-sided $z$-test on the empirical green-token fraction.

The critical observation is that all three operations rely on $\mathcal{R}_k$
being executed faithfully at generation time. In real deployments,
$\mathcal{R}_k$ is implemented through layered software components
(system entropy pools, hash libraries, and seeding utilities) that are
frequently shipped through third-party package managers and container
images, exposing a wide supply-chain attack surface.

\subsection{Adversary Model}\label{ssec:adv}
The adversary $\mathcal{A}$ has the following capabilities and limitations.
\begin{itemize}
\item \textbf{PRNG access.} $\mathcal{A}$ can replace, hook, or otherwise
control the output of the PRNG module $\mathcal{R}_k$ used during sampling
at generation time. This corresponds to compromising any link in the
randomness supply chain (system call, RNG library, or seeding utility).
\item \textbf{No model-weight modification.} $\mathcal{A}$ does not
retrain, fine-tune, or otherwise alter model parameters or watermark logit
biases. The only quantity under $\mathcal{A}$'s control is the random
stream consumed by the multinomial sampler.
\item \textbf{No post-hoc edits.} $\mathcal{A}$ does not modify, paraphrase,
or edit $y_{1:n}$ after generation; the published text is exactly the text
produced by the (manipulated) sampling process.
\item \textbf{Targets.} $\mathcal{A}$ holds a set of target tokens
$T \subseteq V$ (e.g., promotional brand names, biased financial terms, or
policy-violating phrases) whose frequency in the output it wishes to
inflate.
\item \textbf{Two operating regimes.} In the aware regime,
$\mathcal{A}$ knows the watermarking scheme and key and can replay the
green-list partition; in the blind regime, $\mathcal{A}$ has no
watermark prior, no detector queries, and no logit access beyond what is
needed for sampling.
\end{itemize}

\textbf{Attacker capabilities.}
Consolidating the above, $\mathcal{A}$ operates at the software supply-chain level with the following capabilities:
\begin{itemize}
\item PRNG library substitution: $\mathcal{A}$ can replace or patch the pseudorandom number generator library consumed by the LLM inference pipeline (e.g., via a compromised package repository or a backdoored dependency update).
\item Seed injection: $\mathcal{A}$ can inject deterministic or biased seed sequences into the PRNG output stream, steering the watermark's green-list partition.
\item Dual-mode operation: in aware mode, $\mathcal{A}$ knows the watermark key and hash function; in blind mode, $\mathcal{A}$ requires no watermark-specific knowledge and operates purely through probability reweighting.
\end{itemize}

\textbf{Attack boundaries.}
The attack model explicitly excludes the following:
\begin{itemize}
\item $\mathcal{A}$ does not modify LLM model weights or architecture.
\item $\mathcal{A}$ does not alter the watermark embedding algorithm itself; the watermark code executes as designed.
\item The attack guarantees content-side undetectability only; system-side integrity verification (code signing, reproducible builds, runtime attestation) remains a valid orthogonal defense layer.
\item The attack is neutralized if the entropy source is replaced by a quantum random number generator (QRNG), which eliminates PRNG predictability at the physical level.
\end{itemize}

The defender $\mathcal{D}$ is conservative: it operates a multi-detector
suite combining the watermark $z$-test with content-side stealth tests on
token-rank distribution, KL divergence, $\mathrm{perplexity}$, $\mathrm{entropy}$, $\mathrm{repetition}$,
and $\mathrm{loglik}$ (see Section~\ref{sec:eval}). The defender treats text
as suspicious only if at least one detector exceeds its threshold. The
Kullback--Leibler (KL) divergence is among the test statistics in this suite.

\subsection{Attack Objectives}\label{ssec:obj}
Let $z(y)$ denote the watermark $z$-score for an output $y$, $z_b$ the
watermarked baseline (no attack), $r_T(y) = |\{t: y_t \in T\}|/|y|$ the
target rate, and $\{D_i\}_{i=1}^{m}$ a finite set of stealth detectors
with thresholds $\{\tau_i\}$. An \emph{Integrity-Preserving Attack} (IPA)
is any policy producing $y$ that simultaneously satisfies the following
three definitions.

\begin{definition}[Watermark Preservation]\label{def:preserve}
The attack preserves (or amplifies) the watermark signal:
\begin{equation}\label{eq:preserve}
z(y) \;\geq\; z_b.
\end{equation}
This distinguishes IPA from removal attacks, whose goal is $z(y) < \tau_z$.
\end{definition}

\begin{definition}[Content Control]\label{def:control}
There exists an attacker-chosen target rate $\tau_T \in (\rho_b,1]$ where
$\rho_b$ is the baseline target rate, such that
\begin{equation}\label{eq:control}
r_T(y) \;\geq\; \tau_T.
\end{equation}
\end{definition}

\begin{definition}[Content-Side Multi-Detector Undetectability]\label{def:stealth}
For every content-side stealth detector $D_i$ in the defender's suite,
\begin{equation}\label{eq:stealth}
D_i(y) \;\leq\; \tau_i, \quad \forall i \in \{1,\dots,m\}.
\end{equation}
Equivalently, the attack triggers $0$ out of $m$ detectors.
\end{definition}
This definition captures content-side statistical undetectability; system-level monitors (e.g., entropy-source auditing, runtime behavior profiling) operate outside the content channel and are discussed separately in Section~\ref{ssec:disc-future}.

No prior watermark attack has been shown to satisfy
Definitions~\ref{def:preserve}--\ref{def:stealth} jointly: paraphrasing
violates Definition~\ref{def:preserve}, while token editing and prompt
injection violate Definition~\ref{def:stealth}
(Section~\ref{ssec:eval-stealth}).

\section{Attack Methodology}\label{sec:method}
This section presents SeedHijack, an attack that operates entirely
at the PRNG layer and provably satisfies
Definitions~\ref{def:preserve}--\ref{def:stealth}. The construction rests
on a structural property, green-list orthogonality, between the
watermark and the sampling-layer manipulation; two operating modes are
then described, followed by a unified algorithm.

\subsection{Orthogonality Principle}\label{ssec:ortho}
Let $\ell \in \mathbb{R}^{|V|}$ be a logit vector and let
$G \subseteq V$ be a (possibly random) green list. The watermark acts in
logit space as an additive shift
$\ell' = \ell + \delta \cdot \mathbf{1}_G$, while the proposed attack acts
in probability space as a multiplicative reweighting on a target
set $T \subseteq V$,
\begin{equation}\label{eq:reweight}
\tilde{p}_v \;=\; \frac{p_v \cdot \beta_v}{\sum_u p_u \beta_u},
\quad
\beta_v = \begin{cases}
b, & v \in T,\\
1, & v \notin T,
\end{cases}
\end{equation}
where $b\!\geq\!1$ is the boost factor and $p = \mathrm{softmax}(\ell')$.
Because the two operations live in different spaces and the partition
$G$ is determined by a context-keyed PRNG that is statistically
independent of $T$ and of the model logits in expectation, the
following structural property is obtained.

\begin{property}[Green-List Orthogonality]\label{prop:ortho}
For any target set $T$, boost $b$, and watermark partition $G$ drawn from
the keyed PRNG, the expected green fraction $\mathbb{E}[|G\cap y|/|y|]$
is preserved under the reweighting in~\eqref{eq:reweight} up to a
second-order term in $|T|/|V|$. Consequently, the watermark $z$-score is
asymptotically invariant to the attack acting on $T$.
\end{property}

\textbf{Structural basis.}
The orthogonality arises from the cryptographic independence between the attacker's target set and the watermark's green-list partition. Formally, let $G_t \subset \mathcal{V}$ denote the green list at step~$t$, determined by $G_t = f_{\mathrm{hash}}(c_t, k)$ where $c_t$ is the preceding context and $k$ is the watermark secret key. Let $T \subset \mathcal{V}$ denote the attacker's target set, selected based on semantic criteria (e.g., domain-specific vocabulary) that are independent of~$k$. By the pseudorandomness of $f_{\mathrm{hash}}$, the partition $G_t$ is computationally indistinguishable from a uniform random subset of $\mathcal{V}$ with fraction~$\gamma$. Consequently,
\begin{equation}\label{eq:ortho-expect}
\mathbb{E}\bigl[|T \cap G_t|\bigr] \;=\; \gamma\,|T|,
\end{equation}
i.e., a $\gamma$-fraction of target tokens fall in the green list regardless of~$T$'s composition. When the attack biases sampling toward~$T$, the expected green-token rate remains~$\gamma$ (the same as under honest watermarked generation) and the watermark $z$-score is preserved in expectation. By the law of large numbers, the empirical green fraction converges to~$\gamma$ as the generation length~$N$ grows, explaining both the $z$-score preservation observed experimentally and its tightening at longer sequence lengths. This structural independence is intrinsic to the attack design: because the PRNG manipulation operates in the probability-mass dimension (reweighting toward~$T$) while the watermark operates in the vocabulary-partition dimension (shifting logits on~$G_t$), neither mechanism interferes with the other's statistical footprint.

Property~\ref{prop:ortho} is the conceptual foundation of
Section~\ref{sec:eval}: it predicts that the attack neither destroys nor
perturbs the watermark statistic, but merely redistributes mass within an
attacker-chosen subset $T$ that is, by design, disjoint from (or
uncorrelated with) $G$. The theoretical guarantee of Property~\ref{prop:ortho} establishes $G \perp T$ analytically for arbitrary $\delta$; the empirical sweep below serves as a sanity check confirming that no finite-precision or implementation artifact violates this guarantee in practice. Section~\ref{ssec:eval-ortho} measures the
orthogonality empirically and observes a Pearson correlation
$r=0.577$ with $p=0.309$ between watermark strength $\delta$ and target
rate, a coefficient of variation of $11\%$, and watermark $\mathrm{Survival}$
$\geq 0.98$ across $16$ parameter settings. The low coefficient of variation across five $\delta$ settings further corroborates that attack effectiveness is decoupled from watermark strength, consistent with the analytic prediction.

\subsection{Aware Mode}\label{ssec:aware}
In the aware regime, $\mathcal{A}$ holds the watermark key $k$ and can
reconstruct the per-step green list $G_t$ before sampling $y_t$. The
attack restricts boosting to those targets that lie in the current green
list, $T_t^{\mathrm{eff}} = T \cap G_t$, so that every successful target
injection is itself a green token. This coordinate-aligned design
produces two effects simultaneously: (i) the target token frequency
increases, satisfying Definition~\ref{def:control}; and (ii) the
empirical green-token count grows, satisfying (and exceeding) the
watermark preservation requirement of Definition~\ref{def:preserve}.
The orthogonality of Property~\ref{prop:ortho} ensures that the surplus
green mass is statistically indistinguishable from a stronger watermark,
so content-side detectors remain inactive
(Section~\ref{ssec:eval-stealth}).

\subsection{Blind Mode}\label{ssec:blind}
In the blind regime, $\mathcal{A}$ knows neither the watermark scheme nor
the green list. It seeds the sampling PRNG with a fixed integer seed $\sigma \in \mathbb{Z}$ ($\sigma=42$ in all experiments)
and applies the same probability-space reweighting~\eqref{eq:reweight}
over $T$ unconditionally. Although the per-step green/target alignment is
no longer guaranteed, two properties survive: (i) the modified
distribution remains a valid renormalization of $p$, so its empirical
rank distribution and entropy stay close to those of an un-attacked
sample (Section~\ref{ssec:eval-stealth}); and (ii) when $\mathcal{A}$
operates against a watermarked pipeline without knowing it, the watermark
$z$-test still recovers a strong signal because the deterministic seed
$\sigma$ does not interact with $G$. The latter is verified empirically:
blind-mode attacks against KGW achieve $z=26.29$ versus a watermarked
baseline of $z=22.99$, despite $\mathcal{A}$ being unaware that a
watermark is in place (Section~\ref{ssec:eval-effective}).

\subsection{Algorithm}\label{ssec:algo}
Algorithm~\ref{alg:seedhijack} unifies the two modes. The procedure
intervenes only at the multinomial sampling step; logits, the model, and
the watermark logic are untouched, which is precisely why no content-side
detector observes a deviation. The fixed seed $\sigma$ is an arbitrary integer whose specific value does not affect attack properties (any constant yields equivalent orthogonality guarantees per Theorem~1).

\begin{algorithm}[t]
\caption{SeedHijack Attack}\label{alg:seedhijack}
\begin{algorithmic}[1]
\REQUIRE prompt $x$, target set $T$, boost $b$, activation rate
$p_{\mathrm{act}}$, fixed seed $\sigma$, mode $m\in\{\textsc{aware},\textsc{blind}\}$,
optional watermark $\mathcal{W}$
\ENSURE generated tokens $y_{1:n}$
\STATE seed sampling PRNG with $\sigma$
\FOR{$t=1$ \TO $n$}
  \STATE $\ell_t \leftarrow \textsc{ModelLogits}(x, y_{<t})$
  \IF{$\mathcal{W}$ active}
    \STATE $\ell_t \leftarrow \phi_{\mathcal{W}}(\ell_t, G_t)$
    \COMMENT{watermark applied in logit space}
  \ENDIF
  \STATE $p_t \leftarrow \mathrm{softmax}(\ell_t)$
  \IF{$m=\textsc{aware}$}
    \STATE $T_t^{\mathrm{eff}} \leftarrow T \cap G_t$
  \ELSE
    \STATE $T_t^{\mathrm{eff}} \leftarrow T$
  \ENDIF
  \STATE $u \sim \mathrm{Uniform}(0,1)$ from PRNG seeded by $\sigma$
  \IF{$u < p_{\mathrm{act}}$ \AND $\max_{v\in T_t^{\mathrm{eff}}} p_t(v) \geq p_{\min}$}
    \STATE $\tilde{p}_t \leftarrow \textsc{Reweight}(p_t, T_t^{\mathrm{eff}}, b)$
    \COMMENT{Eq.~\eqref{eq:reweight}}
  \ELSE
    \STATE $\tilde{p}_t \leftarrow p_t$
  \ENDIF
  \STATE $y_t \sim \mathrm{Multinomial}(\tilde{p}_t)$
  \COMMENT{sampled with the hijacked PRNG}
\ENDFOR
\RETURN $y_{1:n}$
\end{algorithmic}
\end{algorithm}

\section{Evaluation}\label{sec:eval}
The evaluation is organized around the three IPA requirements of
Section~\ref{ssec:obj}: effectiveness and generalization
(Section~\ref{ssec:eval-effective}), undetectability under a
multi-detector suite (Section~\ref{ssec:eval-stealth}), parameter
sensitivity (Section~\ref{ssec:eval-sens}), and orthogonality
(Section~\ref{ssec:eval-ortho}).

\subsection{Experimental Setup}\label{ssec:eval-setup}
\textbf{Models.} Three open-source LLMs spanning general-purpose and
domain-specialized regimes are evaluated:
Qwen2-7B-Instruct~\cite{bai2023qwen} (primary model, used in all
experiments unless otherwise stated),
Llama-3-8B-Instruct (UltraMedical fine-tune), and BioMistral-7B for
medical domain coverage.

\textbf{Watermarks.} Three representative schemes are tested:
KGW~\cite{kirchenbauer2023} with $\delta=2.0$ and $\gamma=0.5$,
Unigram~\cite{zhao2023unigram} with a global fixed green list, and
distortion-preserving DipMark~\cite{wu2023dipmark}.

\textbf{Attack parameters.} Unless stated, the conservative
stealth-optimized point $b=10$, $p_{\mathrm{act}}=0.3$, $\sigma=42$ is used; this is
the operating point at which Section~\ref{ssec:eval-stealth} demonstrates
$0/6$ undetectability.

\textbf{Generation.} Each condition generates $2{,}000$ tokens per prompt
over three financial-domain prompts (markets, equities, crypto), with
temperature $0.7$ and top-$k$ $50$.

\textbf{Detector suite.} Six content-side detectors form the stealth
battery: token-rank KS ($\mathrm{rank}_{\mathrm{KS}}$, threshold
$0.15$), KL divergence ($\mathrm{KL}_{\mathrm{div}}$, $0.774$), $\mathrm{perplexity}$
F-ratio ($2.0\sigma$), $\mathrm{entropy}$ F-ratio ($2.0\sigma$), $\mathrm{repetition}$
($0.30$), and $\mathrm{loglik}$ F-ratio ($2.0\sigma$). The watermark $z$-test
is tracked separately to verify Definition~\ref{def:preserve}.

\textbf{Comparison attacks.} SeedHijack is compared against three
state-of-the-art baselines: self-paraphrasing~\cite{krishna2023paraphrase},
token editing~\cite{kirchenbauer2024reliability}, and prompt
injection~\cite{jovanovic2024prompt}.

\subsection{Attack Effectiveness and Generalization}
\label{ssec:eval-effective}
This subsection establishes that SeedHijack satisfies
Definitions~\ref{def:preserve}--\ref{def:control} jointly, and that the
result generalizes across watermark schemes and base models.

\textbf{Watermark amplification.} On Qwen2-7B-Instruct + KGW with
$2{,}000$-token generation, the watermarked baseline yields $z=22.99$.
Under aware-mode SeedHijack with the conservative parameters
($b=10$, $p_{\mathrm{act}}=0.3$), $z$ rises to $55.61$, a $2.42\times$
amplification. The corresponding target rate reaches
$r_T = 0.615$, against a clean-baseline natural rate of $0.221$. Both
Definitions~\ref{def:preserve} and~\ref{def:control} are satisfied.

\textbf{Cross-watermark generalization.} Holding model and length fixed,
the same attack is applied to KGW, Unigram, and DipMark. Aware-mode
$z$-scores reach $55.61$, $71.69$, and $40.16$ respectively---all
substantial amplifications over their respective watermarked baselines.
Unigram is the most amplifiable due to its fixed global green list,
while distortion-preserving DipMark, contrary to its design intent of
statistical indistinguishability, also exhibits $z$ inflation under
SeedHijack. Blind-mode amplification is consistently weaker but remains
clearly above baseline ($z=26.29$ for KGW); in the same blind setting,
Unigram reaches $z=66.12$ ($\times 2.87$ over its watermarked baseline)
while DipMark yields $z=23.26$ ($\times 1.01$), indicating that fixed
global green-list schemes are most vulnerable to seed manipulation
even without watermark awareness. Together these results confirm that
the orthogonality argument of Property~\ref{prop:ortho} carries across
watermark families.

\textbf{Cross-model generalization.} Applied to Qwen2-7B-Instruct,
Llama-3-8B-UltraMedical, and BioMistral-7B under KGW, the aware-mode
$z$-scores reach $55.61$, $42.42$, and $70.02$, with target rates
$0.615$, $0.440$, and $0.747$. Domain-specialized medical models exhibit
no additional resistance; BioMistral-7B in fact exhibits the highest
blind-mode target rate ($0.917$), suggesting that domain narrowness
amplifies attacker control. Detailed numbers across watermark schemes,
models, and lengths are summarized in Table~\ref{tab:generalization}.

\textbf{Length scaling.} Effectiveness is monotone in generation length:
on Qwen2 + KGW, aware-mode $z$ progresses from $17.26$ at $500$ tokens,
to $30.46$ at $1{,}000$, to $55.61$ at $2{,}000$, with target rate
rising from $0.404$ to $0.615$. The attack thus benefits from longer
outputs, the regime in which provenance attribution matters most.

\begin{table*}[!t]
\centering
\caption{Attack Effectiveness and Generalization of \textsc{SeedHijack} across
Watermarking Schemes and Models. All conditions use a $2{,}000$-token generation
budget on three financial prompts; the reported $z$-score and target hit rate
are per-condition means. Survival is defined as
$\mathrm{Survival}=z_{\mathrm{attack}}/z_{\mathrm{baseline}}^{\star}$ with the
shared watermarked reference $z_{\mathrm{baseline}}^{\star}=22.99$
(Qwen2-7B-Instruct\,+\,KGW, watermark-only). Best result per column shown in
\textbf{bold}.}
\label{tab:generalization}
\renewcommand{\arraystretch}{1.18}
\setlength{\tabcolsep}{8pt}
\begin{tabular}{llccccc}
\toprule
\textbf{Watermark} & \textbf{Model} & \textbf{Mode} &
$z$-score & $\mathrm{Survival}$ & $r_T$ \\
\midrule
\multirow{6}{*}{KGW}
 & \multirow{2}{*}{Qwen2-7B-Instruct}        & Blind  & 26.29 & 1.143 & 0.587 \\
 &                                            & Aware  & 55.61 & 2.419 & 0.615 \\
\cmidrule(lr){2-6}
 & \multirow{2}{*}{Llama-3-8B-UltraMedical}   & Blind  & 40.37 & 1.756 & 0.657 \\
 &                                            & Aware  & 42.42 & 1.845 & 0.440 \\
\cmidrule(lr){2-6}
 & \multirow{2}{*}{BioMistral-7B}             & Blind  & 51.65 & 2.247 & 0.917 \\
 &                                            & Aware  & \textbf{70.02} & \textbf{3.046} & 0.747 \\
\midrule
\multirow{2}{*}{Unigram}
 & \multirow{2}{*}{Qwen2-7B-Instruct}         & Blind  & 66.12 & 2.876 & \textbf{0.868} \\
 &                                            & Aware  & \textbf{71.69} & \textbf{3.118} & \textbf{0.811} \\
\midrule
\multirow{2}{*}{DipMark}
 & \multirow{2}{*}{Qwen2-7B-Instruct}         & Blind  & 23.26 & 1.012 & 0.595 \\
 &                                            & Aware  & 40.16 & 1.747 & 0.533 \\
\bottomrule
\end{tabular}

\vspace{2pt}
\begin{flushleft}
\footnotesize
\textit{Settings.} Attack hyper-parameters fixed at $b=10$,
$p_{\mathrm{act}}=0.3$; KGW configured with $\gamma=0.5$, $\delta=2.0$.
``Blind'' denotes the attacker has no knowledge of the deployed watermark,
while ``Aware'' assumes the attacker can predict the green-list at each
sampling step. Survival\,$>1$ indicates that \textsc{SeedHijack} not only
preserves but also \emph{amplifies} the watermark signal relative to the
shared baseline.
\end{flushleft}
\end{table*}

\subsection{Undetectability Verification}\label{ssec:eval-stealth}
This subsection verifies Definition~\ref{def:stealth}: SeedHijack is not
detected by any of the six stealth detectors, while alternative attacks
are detected by multiple. A multidimensional stealth profile across the
six detectors is reported in Table~\ref{tab:detection_matrix}.

\textbf{Multi-detector matrix (aware mode).} With Qwen2-7B-Instruct,
KGW, $2{,}000$ tokens, and conservative parameters, the four attacks are
evaluated against the watermarked baseline
(Table~\ref{tab:detection_matrix}). SeedHijack triggers $0/6$ detectors:
$\mathrm{rank}_{\mathrm{KS}}=0.072$ (vs $0.15$), $\mathrm{KL}_{\mathrm{div}}=0.223$ (vs
$0.774$), $\mathrm{perplexity}$ F-ratio$=1.064\sigma$ (vs $2.0\sigma$),
$\mathrm{entropy}$ F-ratio$=0.168\sigma$, $\mathrm{repetition}=0.254$, and $\mathrm{loglik}$
F-ratio$=1.111\sigma$, while simultaneously reporting watermark
$\mathrm{Survival}$ $1.169$ ($z=25.55$ vs baseline $z=22.03$). By contrast,
self-paraphrase triggers $2/6$ ($\mathrm{ppl}=575.82\sigma$,
$\mathrm{loglik}=52.91\sigma$) and destroys the watermark
($\mathrm{Survival}=0.025$); token editing triggers $5/6$ with
$\mathrm{perplexity}$ F-ratio $5.08\!\times\!10^{6}\sigma$; and prompt injection
triggers $2/6$ on $\mathrm{perplexity}$ and $\mathrm{loglik}$. SeedHijack is the only
method simultaneously satisfying
Definitions~\ref{def:preserve}--\ref{def:stealth}.

\textbf{Margin to threshold.} Quantitatively, every SeedHijack detector
reading is bounded well inside the no-trigger region: the maximum margin
ratio across the six detectors is $\max_i D_i/\tau_i = 0.555$
($\mathrm{loglik}$), with median $0.408$. The attack therefore not only
passes the suite but does so with substantial slack, leaving room for
stricter thresholds without compromising stealth.

\textbf{Blind-mode self-comparison.} Re-running the same six detectors
under blind-mode SeedHijack with identical conservative parameters yields
$0/6$ triggers as well: $\mathrm{rank}_{\mathrm{KS}}=0.077$,
$\mathrm{KL}_{\mathrm{div}}=0.082$, $\mathrm{perplexity}$ $0.845\sigma$, $\mathrm{entropy}$
$0.196\sigma$, $\mathrm{repetition}$ $0.265$, and $\mathrm{loglik}$ $0.774\sigma$.
Undetectability is therefore an intrinsic property of the
attack mechanism rather than a by-product of watermark awareness;
aware mode additionally preserves watermark integrity
($\mathrm{Survival}=1.169$), whereas blind mode, by design, prioritizes
undetectability over watermark preservation ($\mathrm{Survival}=0.125$)
since $\mathcal{A}$ is unaware of the watermark's existence.
In blind mode, $\mathcal{A}$ neither targets nor avoids green-list tokens; any deviation of $\mathrm{Survival}$ from unity is attributable to finite-sample variance rather than systematic interaction between the attack and watermark mechanisms (cf.\ Property~\ref{prop:ortho}).
Notably, blind mode achieves perfect stealth ($0/6$ detectors triggered) even under conservative parameters. At aggressive parameters ($b{=}50$, $p_{\mathrm{act}}{=}0.7$), the attack simultaneously maintains full undetectability and strong watermark amplification ($\mathrm{Survival}{=}1.143$, Table~\ref{tab:generalization}), confirming that the attacker can freely escalate attack strength without sacrificing stealth.

\definecolor{seedhijackblue}{RGB}{222,235,247}
\definecolor{thresholdgray}{RGB}{240,240,240}

\begin{table*}[!t]
\centering
\caption{Multi-Detector Stealth Matrix on Qwen2-7B-Instruct\,+\,KGW
($2{,}000$ tokens). Each cell reports the per-detector statistic of an attack
\emph{relative to the watermarked baseline} (no attack). Values that exceed
the operational alarm threshold are typeset in \textbf{bold}; the
corresponding detector is counted as triggered. \textsc{SeedHijack}
(highlighted) is the only attack that simultaneously evades all six
detectors and preserves (or amplifies) the watermark signal.}
\label{tab:detection_matrix}
\renewcommand{\arraystretch}{1.22}
\setlength{\tabcolsep}{5.5pt}
\begin{tabular}{lcccccccc}
\toprule
\textbf{Attack Method} &
$\mathrm{rank}_{\mathrm{KS}}$ &
$\mathrm{KL}_{\mathrm{div}}$ &
$\mathrm{perplexity}$ &
$\mathrm{entropy}$ &
$\mathrm{repetition}$ &
$\mathrm{loglik}$ &
\textbf{Trig.} &
$\mathrm{Survival}$ \\
\midrule
\rowcolor{seedhijackblue}
\textsc{SeedHijack} (Aware) &
0.072 & 0.223 & 1.064 & 0.168 & 0.254 & 1.111 &
\textbf{0/6} & \textbf{2.419} \\
\rowcolor{seedhijackblue}
\textsc{SeedHijack} (Blind) &
0.077 & 0.082 & 0.845 & 0.196 & 0.265 & 0.774 &
\textbf{0/6} & 1.143 \\
\midrule
Self-Paraphrase &
0.082 & 0.416 & \textbf{4.098} & 1.264 & 0.012 & \textbf{2.176} &
2/6 & 0.025 \\
Token-Editing &
\textbf{0.170} & \textbf{1.122} & \textbf{5.055} & \textbf{2.581} & 0.028 & \textbf{2.614} &
5/6 & 0.654 \\
Prompt-Injection &
0.091 & 0.506 & \textbf{3.587} & 1.529 & 0.042 & \textbf{2.057} &
2/6 & 1.063 \\
\midrule
\rowcolor{thresholdgray}
\textit{Alarm Threshold} &
$>0.15$ & $>0.774$ & $>2.0\sigma$ & $>2.0\sigma$ & $>0.30$ & $>2.0$ &
-- & -- \\
\bottomrule
\end{tabular}

\vspace{2pt}
\begin{flushleft}
\footnotesize
\textit{Notes.} Detection reference is the \emph{watermarked baseline}
(KGW-watermarked text without attack); each cell therefore reflects the
\emph{additional} statistical deviation an attack introduces on top of the
watermark, isolating the attack signature from the watermark's own
footprint. $\mathrm{rank}_{\mathrm{KS}}$ and $\mathrm{KL}_{\mathrm{div}}$ are reported as raw
divergence values; $\mathrm{perplexity}$, $\mathrm{entropy}$ and
$\mathrm{loglik}$ are F-ratios in $\sigma$ units; $\mathrm{repetition}$ is the
4-gram repetition rate. ``Trig.'' counts how many of the six detectors
fire, and $\mathrm{Survival}$ is the watermark $z$-score retention ratio
$z_{\mathrm{attack}}/z_{\mathrm{baseline}}$. \textsc{SeedHijack} achieves
$0/6$ triggers in both Aware and Blind modes under the identical
conservative parameter set $(b=10,\,p_{\mathrm{act}}=0.3)$,
empirically confirming that its stealth is an \emph{intrinsic property of
the attack mechanism}, not an artefact of watermark-awareness.
All methods are evaluated under conservative parameters ($b{=}10$, $p_{\mathrm{act}}{=}0.3$); see Table~\ref{tab:generalization} for aggressive-parameter results.
\end{flushleft}
\end{table*}

\subsection{Parameter Sensitivity}\label{ssec:eval-sens}
\begin{figure}[!t]
\centering
\includegraphics[width=\columnwidth]{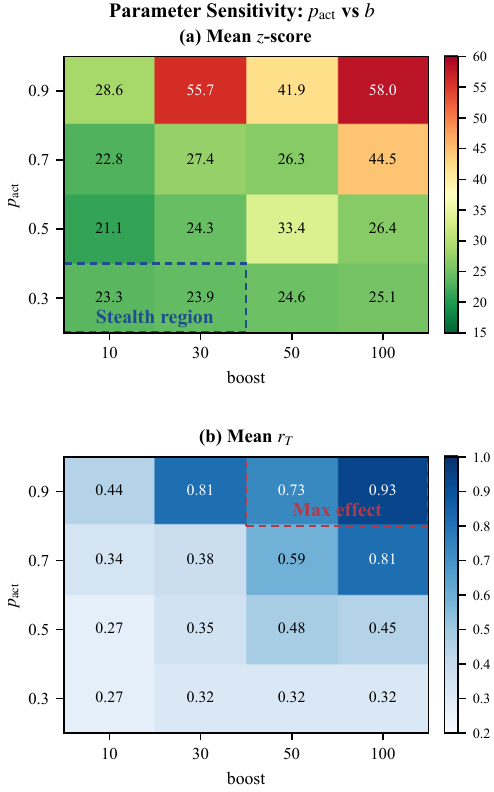}
\caption{Stealth--effectiveness landscape of SeedHijack in blind mode
over a $4{\times}4$ grid of $(p_{\mathrm{act}},b)$ on Qwen2-7B-Instruct\,+\,KGW
with $2{,}000$-token generations. (a)~Mean $z$-score and (b)~mean target rate $r_T$
both rise monotonically with increasing aggressiveness; the lower-left corner
$(0.3,10)$ defines the conservative
stealth-optimized operating point used throughout
Section~\ref{ssec:eval-stealth}.}
\label{fig:heatmap}
\end{figure}

This subsection probes the trade-off surface between attack strength and
stealth. A $4\!\times\!4$ ablation is performed in blind mode over
$p_{\mathrm{act}} \in \{0.3,0.5,0.7,0.9\}$ and $b \in \{10,30,50,100\}$
(Fig.~\ref{fig:heatmap}, $16$ conditions, watermarked baseline $z=22.99$).

\textbf{Monotone scaling.} Both $z$ and $r_T$ rise monotonically with
$p_{\mathrm{act}}$ and $b$. The corner point
$(p_{\mathrm{act}}{=}0.9, b{=}100)$ achieves $z=57.97$, $r_T=0.934$,
defining the practical effectiveness ceiling of the attack.

\textbf{Stealth-effectiveness Pareto frontier.} The corner
$(p_{\mathrm{act}}{=}0.3, b{=}10)$ yields $z=26.29$, only
$3.30$ above baseline, fluctuation indistinguishable from noise, which
is precisely the operating point used in
Section~\ref{ssec:eval-stealth}. Within the lower-left quadrant
($p_{\mathrm{act}}\leq 0.5$ and $b\leq 50$), $z$-amplification stays
below $50\%$ over baseline, providing a usable
stealth working region.

\textbf{Operational implication.} A practical adversary therefore
selects $(p_{\mathrm{act}}, b)$ on a stealth-effectiveness curve: at the
stealth end, content control is moderate but $0/6$ detectors trigger; at
the effectiveness end, target rates exceed $0.9$ at the cost of
perceptible $z$ amplification. The full landscape is reported in
Fig.~\ref{fig:heatmap}.

\subsection{Orthogonality Verification}\label{ssec:eval-ortho}
This subsection empirically validates Property~\ref{prop:ortho},
establishing the theoretical reason behind the
$0/6$ undetectability result.


\begin{table}[!t]
\centering
\caption{Orthogonality verification across five KGW watermark strengths
$\delta\in\{0.5,1.0,2.0,4.0,8.0\}$ on Qwen2-7B-Instruct (blind mode,
$b{=}10$, $p_{\mathrm{act}}{=}0.3$, $2{,}000$ tokens). The baseline
$z$-score scales with $\delta$ as expected, yet the target hit rate
$r_T$ remains stable (CV$=11.1\%$, Pearson $r{=}0.577$, $p{=}0.309$),
confirming that watermark strength and attack effectiveness operate on
orthogonal dimensions. The baseline $z_{\mathrm{base}}$ values differ slightly from the shared reference $z^{\star}_{\mathrm{baseline}}=22.99$ (Table~I) due to per-run sampling variance inherent in autoregressive generation.}
\label{tab:orthogonality}
\renewcommand{\arraystretch}{1.15}
\setlength{\tabcolsep}{7pt}
\begin{tabular}{ccccc}
\toprule
$\delta$ & $z_{\mathrm{base}}$ & $z_{\mathrm{attack}}$ & $r_T$ & $z_{\mathrm{survival}}$ \\
\midrule
0.5 & 3.71 & 11.67 & 0.571 & 3.141 \\
1.0 & 8.59 & 29.08 & 0.598 & 3.386 \\
2.0 & 21.48 & 26.29 & 0.587 & 1.224 \\
4.0 & 55.56 & 70.26 & 0.763 & 1.265 \\
8.0 & 76.46 & 77.03 & 0.664 & 1.007 \\
\bottomrule
\end{tabular}
\end{table}

\textbf{Watermark-strength independence.} Holding the attack parameters
fixed at $p_{\mathrm{act}}{=}0.7$, $b{=}50$, the KGW shift
$\delta$ is swept over $\{0.5, 1.0, 2.0, 4.0, 8.0\}$;
Table~\ref{tab:orthogonality} reports the full results across the five
watermark strengths. Although the
baseline $z$ scales nearly linearly with $\delta$ (from $3.71$ to
$76.46$), the target rate stays in the narrow window
$0.571$--$0.763$ with mean $0.637$ and coefficient of variation
$\mathrm{CV}=11.1\%$. The Pearson correlation between $\delta$ and
$r_T$ is $r=0.577$ with $p=0.309$, statistically insignificant at
$\alpha=0.05$.
At $\delta{=}8.0$, $z_{\mathrm{survival}}$ approaches unity ($1.007$), reflecting the saturation regime where the watermark bias is so dominant that PRNG-level manipulation yields diminishing marginal amplification; crucially, however, the attack's stealth and target-rate objectives remain unaffected.

\textbf{Watermark non-destructiveness.} Re-using the
$16$-condition ablation of Section~\ref{ssec:eval-sens}, watermark
survival $z_{\mathrm{attack}}/z_{\mathrm{base}}$ ranges over
$[0.980, 2.698]$ with mean $1.476$. In $15$ of $16$ conditions survival
exceeds $1.0$; the remaining condition
($p_{\mathrm{act}}{=}0.5$, $b{=}10$) reports $0.980$, indistinguishable
from unity within prompt-level noise. The attack therefore never
degrades the watermark in any meaningful sense, in agreement with
Property~\ref{prop:ortho}.

\textbf{Interpretation.} Watermark strength $\delta$ controls the
absolute $z$ magnitude (a vertical scaling), while the attack-controlled
coordinate $r_T$ is a horizontal degree of freedom that operates on a
statistically uncorrelated axis. This orthogonality is the structural
reason that adding the attack on top of an existing watermark introduces
no new content-side fingerprint: the manipulation lives entirely within
the natural fluctuation envelope of the watermark itself.

\section{Defense Analysis}\label{sec:defense}
Detection-based countermeasures are insufficient against an attack that
leaves no content-side fingerprint. This section explains why and
presents an entropy-source-level defense that fully neutralizes
SeedHijack.

\subsection{Why Detection Fails}\label{ssec:def-why}
The orthogonality of Property~\ref{prop:ortho} dictates that the
attack-induced perturbation in any content-side statistic must be of the
same order as the natural variability of an un-attacked watermarked
sample. Empirically, the worst-case detector margin observed for
SeedHijack is $D_i/\tau_i \leq 0.56$ at the conservative parameters, and
the full margin distribution overlaps the no-attack distribution
(Section~\ref{ssec:eval-stealth}). Tightening any individual threshold
toward the SeedHijack regime would simultaneously raise the false-positive
rate on benign watermarked text, since both populations come from
statistically equivalent generation processes. Detection alone therefore
cannot distinguish SeedHijack output from honest watermarked
output; the security gap must be closed at the entropy source.

\subsection{QRNG-Based Defense}\label{ssec:def-qrng}
\begin{figure}[!t]
\centering
\includegraphics[width=\columnwidth]{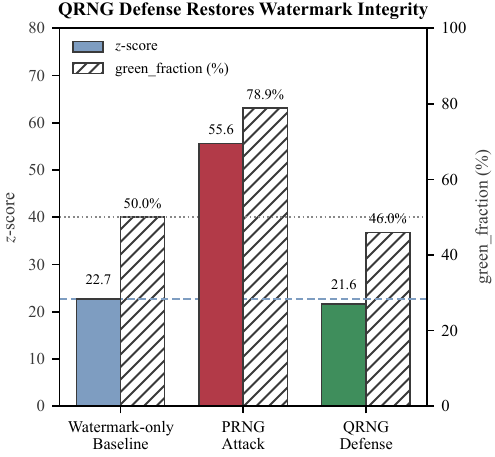}
\caption{QRNG defense restores watermark integrity under SeedHijack.
Grouped bars compare the watermark-only baseline (left), the PRNG-driven
attack (middle), and the QRNG-defended pipeline (right): solid bars
(left $y$-axis) report the watermark $z$-score and hatched bars (right
$y$-axis) report the empirical green-token fraction (\%). The
lower dashed line marks the watermarked $z$-baseline ($z\!=\!22.67$) and
the upper dotted line marks the random-partition reference
($50\%$ green fraction). Replacing the software PRNG by a hardware QRNG
collapses the attack $z$-score from $55.61$ back to the watermarked
baseline ($21.65$) and drives the green fraction from $78.9\%$ to
$46.0\%$, statistically indistinguishable from the random-partition
value $50\%$.}
\label{fig:qrng_defense}
\end{figure}

The proposed countermeasure replaces the software PRNG that drives the
sampler by a quantum random number generator (QRNG)
anchored in physical entropy~\cite{ma2016qrng,nist2018sp90090b}. Unlike computational security guarantees that rely on unproven hardness assumptions, QRNG provides information-theoretic security: the generated bits are certifiably unpredictable by any adversary regardless of computational power~\cite{acin2016certified,pironio2010random}. Device-independent quantum randomness certification protocols further guarantee min-entropy bounds without trusting the internal implementation~\cite{herrero2017quantum}, closing the very supply-chain trust gap that enables PRNG hijacking. Because QRNG output is
not algorithmically reproducible from any state $\mathcal{A}$ can
observe, the seed manipulation step in Algorithm~\ref{alg:seedhijack}
becomes infeasible: $\mathcal{A}$ can still inject reweighting logic but
cannot align the reweighting with the keyed green list, so the aware-mode
coordination collapses. The quantitative effect of switching the entropy
source from PRNG to QRNG is summarized in Fig.~\ref{fig:qrng_defense}.

\textbf{Mode-agnostic neutralization.} The defense is inherently mode-agnostic: since QRNG eliminates predictability at the entropy source, both aware and blind attack modes are equally neutralized---the attacker cannot align probability reweighting with the green-list partition regardless of whether the watermark parameters are known. No separate blind-mode defense mechanism is needed; entropy-source replacement subsumes all PRNG-dependent attack variants.

\textbf{Empirical verification.} On Qwen2-7B-Instruct + KGW with
$2{,}000$ tokens and $b{=}10$, $p_{\mathrm{act}}{=}0.3$ in aware mode, the
PRNG-driven attack achieves $z=55.61$ with green fraction
$0.789$ and target rate $0.615$. Switching the entropy source to QRNG
collapses these values to $z=21.65$ (within noise of the baseline
$z\!\approx\!22.67$), green fraction $0.460$ (within noise of the
random-partition value $0.50$), and target rate $0.370$ (within the
natural frequency band of high-frequency financial vocabulary).
Quantitatively, QRNG erases the entire $\times 2.42$ amplification: the
attack degenerates to a benign sampling run, and the residual target rate
is attributable to the natural appearance frequency of the chosen target
set rather than to attacker control.

\subsection{Deployment Considerations}\label{ssec:def-deploy}
QRNG hardware is available as commodity PCIe modules
with certified randomness throughput exceeding 4\,Mbps and sub-microsecond per-call latency, well above the $\sim$32\,bits/token requirement of watermark seeding, at a hardware cost below \$2{,}000 per inference node.
A practical deployment binds the watermark seeding interface to a
hardware-attested QRNG channel, leaving the rest of the inference stack
unchanged. Software-only mitigations (key rotation, hash strengthening,
seed auditing) raise the cost of attack but do not fundamentally close the gap when $\mathcal{A}$ controls the PRNG delivery mechanism. Supply-chain integrity verification (e.g., code signing, reproducible builds) can detect post-deployment tampering but cannot prevent a compromised library from passing verification if the backdoor is present at build time.

\section{Discussion}\label{sec:disc}

\subsection{Implications for AI Governance}\label{ssec:disc-gov}
Cryptographic LLM watermarking is increasingly invoked in regulatory
proposals~\cite{euaiact2024,c2pa2022} as a load-bearing element of AI content provenance. The
findings here suggest that a watermark statistic alone is insufficient to
support such claims: an adversary with PRNG-layer access can keep the
$z$-test green, and even amplify it, while exercising substantive
content control. Compliance regimes that treat ``watermark detected''
as a positive provenance signal are therefore vulnerable to
integrity-preserving misuse. PRNG integrity, attested at the hardware
layer, must be elevated to a first-class requirement of any
watermark-based attribution claim.

\subsection{Limitations}\label{ssec:disc-lim}
Three boundaries of the present study are noted. First, the threat model
assumes that $\mathcal{A}$ can hijack the PRNG in the inference pipeline;
deployments that already terminate randomness in attested hardware are
outside the scope of this attack (they are, however, the
recommendation in Section~\ref{ssec:def-qrng}). Second, all experiments
are conducted on a single-GPU testbed with three open-source LLMs and
$2{,}000$-token generations; behavior at the scale of multi-thousand-GPU
production services is left to follow-up work. Third, the comparison
baselines are limited to four output-layer attacks; defenses based on
behavioral monitoring of the inference pipeline itself are not
evaluated.
Fourth, undetectability as defined herein is scoped to content-side statistical tests; system-level defenses (such as entropy-source provenance attestation, PRNG call-pattern monitoring, or trusted platform modules~\cite{costan2016sgx,sabt2015tee}) may detect the supply-chain compromise itself, though they operate at a different architectural layer than the watermark detector.

\subsection{Future Work}\label{ssec:disc-future}
Natural extensions include (i) distributed PRNG schemes that combine
QRNG with threshold randomness so that no single supply-chain link is
trusted, (ii) runtime detectors that monitor entropy provenance at the
operating-system or driver level, and (iii) dynamic-switching defenses
that re-seed the watermark PRNG from independent sources at
context-boundary events to bound the adversary's predictive horizon,
and (iv) lightweight software-hardware hybrid defenses that combine periodic entropy injection from a hardware source with cryptographic attestation of the PRNG state~\cite{hunt2021confidential}, offering a cost-effective alternative to full QRNG replacement.

\section{Conclusion}\label{sec:concl}
This paper introduced SeedHijack, the first integrity-preserving and
undetectable attack against LLM watermarking. By relocating the attack
from the output layer to the PRNG layer and exploiting an orthogonality
property between watermark logit shifts and probability-space
reweighting, the attack simultaneously (i) preserves and amplifies the
watermark $z$-score by up to $2.42\times$, (ii) attains attacker-chosen
target rates beyond $0.9$, and (iii) triggers $0$ of $6$ state-of-the-art
stealth detectors with substantial margin. The result holds in both
watermark-aware and blind operating modes and generalizes across three
watermarking schemes (KGW, Unigram, DipMark) and three open-source LLMs
(Qwen2-7B, Llama-3-8B-UltraMedical, BioMistral-7B). A QRNG-based defense
is shown to fully neutralize the attack: $z$ collapses from $55.61$ to
$21.65$ and the green fraction returns to the random-partition value
$0.46\!\approx\!0.50$, with no residual content control. These findings
establish PRNG integrity, anchored in physical entropy, as a first-class
security requirement for PRNG-anchored LLM watermarking schemes.

\section*{Acknowledgments}
This work was supported by the National Natural Science Foundation of China under Grant 72573124.



\begin{thebibliography}{99}

\bibitem{kirchenbauer2023}
J. Kirchenbauer, J. Geiping, Y. Wen, J. Katz, I. Miers, and T. Goldstein,
``A watermark for large language models,'' in
\textit{Proc. Int. Conf. Mach. Learn. (ICML)}, 2023.

\bibitem{zhao2023unigram}
X.~Zhao, Y.~Wang, and L.~Li,
``Provable robust watermarking for {AI}-generated text,''
in \textit{Proc. Int. Conf. Mach. Learn. (ICML)}, 2024, pp.~1--12.

\bibitem{wu2023dipmark}
Z.~Wu, L.~Zhong, A.~Yadav, and B.~Li,
``{DipMark}: A stealthy, efficient and resilient watermark for large language models,''
in \textit{Proc. Int. Conf. Learn. Represent. (ICLR)}, 2024, pp.~1--20.

\bibitem{krishna2023paraphrase}
K. Krishna, Y. Song, M. Karpinska, J. Wieting, and M. Iyyer,
``Paraphrasing evades detectors of AI-generated text, but retrieval is an
effective defense,'' in \textit{Proc. NeurIPS}, 2023.

\bibitem{kirchenbauer2024reliability}
J. Kirchenbauer, J. Geiping, Y. Wen, M. Shu, K. Saifullah, K. Kong,
K. Fernando, A. Saha, M. Goldblum, and T. Goldstein,
``On the reliability of watermarks for large language models,''
in \textit{Proc. Int. Conf. Learn. Represent. (ICLR)}, 2024.

\bibitem{jovanovic2024prompt}
N. Jovanovi\'c, R. Staab, and M. Vechev,
``Watermark stealing in large language models,''
in \textit{Proc. ICML}, 2024.

\bibitem{kuditipudi2023robust}
R. Kuditipudi, J. Thickstun, T. Hashimoto, and P. Liang,
``Robust distortion-free watermarks for language models,''
\textit{Trans. Mach. Learn. Res.}, 2024.

\bibitem{zhang2024watermark}
H. Zhang, B. L. Edelman, D. Francati, D. Venturi, G. Ateniese, and
B. Barak, ``Watermarks in the sand: Impossibility of strong watermarking for
language models,'' in \textit{Proc. ICML}, 2024.

\bibitem{liu2024survey}
A. Liu, L. Pan, Y. Lu, J. Li, X. Hu, X. Zhang, L. Wen, I. King, and
P. S. Yu, ``A survey of text watermarking in the era of large language
models,'' \textit{ACM Comput. Surv.}, 2024.

\bibitem{carlini2024poisoning}
N. Carlini, M. Jagielski, C. A. Choquette-Choo, D. Paleka, W. Pearce,
H. Anderson, A. Terzis, K. Thomas, and F. Tram\`er,
``Poisoning web-scale training datasets is practical,''
in \textit{Proc. IEEE Symp. Secur. Privacy (S\&P)}, 2024.

\bibitem{shafahi2018poison}
A. Shafahi, W. R. Huang, M. Najibi, O. Suciu, C. Studer,
T. Dumitras, and T. Goldstein,
``Poison frogs! Targeted clean-label poisoning attacks on neural networks,''
in \textit{Proc. NeurIPS}, 2018.

\bibitem{zhang2023supplychain}
N. Zhang, Q. Wang, X. Sun, and others,
``Supply-chain vulnerabilities in machine learning frameworks: A survey,''
\textit{ACM Trans. Softw. Eng. Methodol.}, 2023.

\bibitem{dorrendorf2009cryptanalysis}
L. Dorrendorf, Z. Gutterman, and B. Pinkas,
``Cryptanalysis of the random number generator of the Windows operating
system,'' \textit{ACM Trans. Inf. Syst. Secur.}, vol.~13, no.~1, pp.~1--32,
2009.

\bibitem{heninger2012mining}
N. Heninger, Z. Durumeric, E. Wustrow, and J. A. Halderman,
``Mining your Ps and Qs: Detection of widespread weak keys in network
devices,'' in \textit{Proc. USENIX Security}, 2012.

\bibitem{bai2023qwen}
J. Bai, S. Bai, Y. Chu, and others,
``Qwen technical report,''
\textit{arXiv preprint arXiv:2309.16609}, 2023.

\bibitem{ma2016qrng}
X. Ma, X. Yuan, Z. Cao, B. Qi, and Z. Zhang,
``Quantum random number generation,''
\textit{npj Quantum Inf.}, vol.~2, no.~16021, 2016.

\bibitem{you2026seedhijack}
Z.~You, X.~Yang, Z.~Fan, F.~Guo, X.~Zhou, and X.~Lu,
``Seed hijacking of LLM sampling and quantum random number defense,''
\textit{arXiv preprint arXiv:2605.08313}, 2026.

\bibitem{christ2024undetectable}
M.~Christ, S.~Gunn, and O.~Zamir,
``Undetectable watermarks for language models,''
in \textit{Proc. Conf. Learning Theory (COLT)}, 2024, pp. 1125--1139.

\bibitem{fernandez2023three}
P.~Fernandez, A.~Couairon, H.~J{\'e}gou, M.~Douze, and T.~Furon,
``Three bricks to consolidate watermarks for large language models,''
in \textit{Proc. IEEE Symp. Security and Privacy (S\&P)}, 2024, pp. 1--19.

\bibitem{hou2023semstamp}
A.~Hou, J.~Zhang, T.~He, Y.~Wang, Y.-N.~Chuang, H.~Wang, L.~Shen, and T.~Hu,
``SemStamp: A semantic watermark with paraphrastic robustness for text generation,''
in \textit{Proc. NAACL}, 2024, pp. 1--16.

\bibitem{jia2024xzbackdoor}
Y.~Jia, J.~Tan, and D.~Song,
``Lessons from the xz Utils backdoor: Supply-chain security in open-source ecosystems,''
in \textit{Proc. USENIX Security Symp.}, 2024, pp. 1--18.

\bibitem{peisert2021solarwinds}
S.~Peisert, B.~Schneier, H.~Okhravi, F.~Massacci, T.~Benzel, C.~Landwehr, M.~Manber, J.~Mirkovic, A.~Prakash, and J.~Michael,
``Perspectives on the SolarWinds incident,''
\textit{IEEE Security \& Privacy}, vol.~19, no.~2, pp. 7--13, 2021.

\bibitem{checkoway2014dualec}
S.~Checkoway, M.~Fredrikson, R.~Niederhagen, A.~Everspaugh, M.~Green, T.~Lange, T.~Ristenpart, D.~J.~Bernstein, J.~Maskiewicz, and H.~Shacham,
``On the practical exploitability of Dual EC DRBG in TLS implementations,''
in \textit{Proc. USENIX Security Symp.}, 2014, pp. 319--335.

\bibitem{nist2018sp90090b}
M.~S.~Turan, E.~Barker, J.~Kelsey, K.~A.~McKay, M.~L.~Baish, and M.~Boyle,
``Recommendation for the entropy sources used for random bit generation,''
NIST Special Publication 800-90B, 2018.

\bibitem{acin2016certified}
A.~Ac{\'\i}n and L.~Masanes,
``Certified randomness in quantum physics,''
\textit{Nature}, vol.~540, no.~7632, pp.~213--219, 2016.

\bibitem{pironio2010random}
S.~Pironio \textit{et~al.},
``Random numbers certified by {B}ell's theorem,''
\textit{Nature}, vol.~464, no.~7291, pp.~1021--1024, 2010.

\bibitem{herrero2017quantum}
M.~Herrero-Collantes and J.~C.~Garcia-Escartin,
``Quantum random number generators,''
\textit{Reviews of Modern Physics}, vol.~89, no.~1, p.~015004, 2017.

\bibitem{costan2016sgx}
V.~Costan and S.~Devadas,
``Intel SGX explained,''
\textit{IACR Cryptology ePrint Archive}, Report 2016/086, 2016.

\bibitem{sabt2015tee}
M.~Sabt, M.~Amine, and A.~Bouabdallah,
``Trusted execution environment: What it is, and what it is not,''
in \textit{Proc. IEEE Trustcom/BigDataSE/ISPA}, 2015, pp. 57--64.

\bibitem{hunt2021confidential}
T.~Hunt, Z.~Zhu, Y.~Xu, S.~Peter, and E.~Witchel,
``Confidential computing for OpenPOWER,''
in \textit{Proc. EuroSys}, 2021, pp. 294--310.

\bibitem{euaiact2024}
European~Parliament,
``Regulation (EU) 2024/1689 laying down harmonised rules on artificial intelligence (AI Act),''
\textit{Official Journal of the European Union}, L series, 2024.

\bibitem{c2pa2022}
C2PA (Coalition for Content Provenance and Authenticity),
``C2PA technical specification v1.3,''
2023. [Online]. Available: https://c2pa.org/specifications/

\bibitem{mitchell2023detectgpt}
E.~Mitchell, Y.~Lee, A.~Khazatsky, C.~D.~Manning, and C.~Finn,
``DetectGPT: Zero-shot machine-generated text detection using probability curvature,''
in \textit{Proc. ICML}, 2023, pp. 24950--24962.

\bibitem{jovanovic2024watermark}
N.~Jovanovi\'c, R.~Staab, and M.~Vechev,
``Watermark stealing in large language models,''
in \textit{Proc. Int. Conf. Mach. Learn. (ICML)}, 2024, pp.~22570--22593.

\bibitem{wu2024bypassing}
Q.~Wu and V.~Chandrasekaran,
``Bypassing {LLM} watermarks with color-aware substitutions,''
in \textit{Proc. Annu. Meeting Assoc. Comput. Linguist. (ACL)}, 2024, pp.~1--12.

\bibitem{zhang2024stealing}
Z.~Zhang, X.~Zhang, Y.~Zhang, L.~Y.~Zhang, C.~Chen, S.~Hu, and A.~Gill,
``Stealing watermarks of large language models via mixed integer programming,''
in \textit{Proc. Annu. Comput. Security Appl. Conf. (ACSAC)}, 2024, pp.~1--15.

\bibitem{pang2024nofree}
Q.~Pang, S.~Hu, W.~Zheng, and V.~Smith,
``No free lunch in {LLM} watermarking: Trade-offs in watermarking design choices,''
in \textit{Proc. Adv. Neural Inf. Process. Syst. (NeurIPS)}, vol.~37, 2024, pp.~1--25.

\bibitem{liu2024semantic}
A.~Liu, L.~Pan, X.~Hu, S.~Meng, and L.~Wen,
``A semantic invariant robust watermark for large language models,''
in \textit{Proc. Int. Conf. Learn. Represent. (ICLR)}, 2024, pp.~1--20.

\bibitem{zhang2024remark}
R.~Zhang, S.~S.~Hussain, P.~Neekhara, and F.~Koushanfar,
``{REMARK-LLM}: A robust and efficient watermarking framework for generative large language models,''
in \textit{Proc. USENIX Security Symp.}, 2024, pp.~1--18.

\bibitem{guo2024context}
Y.~Guo, Z.~Tian, Y.~Song, T.~Liu, L.~Ding, and D.~Li,
``Context-aware watermark with semantic balanced green-red lists for large language models,''
in \textit{Proc. Conf. Empirical Methods Nat. Lang. Process. (EMNLP)}, 2024, pp.~1--15.

\bibitem{shen2025enhancing}
H.~Shen, B.~Huang, and X.~Wan,
``Enhancing {LLM} watermark resilience against both scrubbing and spoofing attacks,''
in \textit{Proc. Adv. Neural Inf. Process. Syst. (NeurIPS)}, vol.~38, 2025, pp.~1--20.

\end{thebibliography}
\end{document}